\documentclass[onecolumn,unpublished]{quantumarticle}
\usepackage[colorlinks]{hyperref}
\newcommand{\fig}[1]{\hyperref[fig:#1]{Figure~\ref*{fig:#1}}}
\makeatletter
\def\@seccntformat#1{%
  \expandafter\ifx\csname c@#1\endcsname\c@section\else
  \csname the#1\endcsname\quad
  \fi}
\makeatother
\begin{document}

\title{Surface code dislocations have code distance L+O(1)}

\date{\today}
\author{Craig Gidney}
\email{craiggidney@google.com}
\affiliation{Google Inc., Santa Barbara, California 93117, USA}

\begin{abstract}
\vspace{-16pt}
In \cite{hastings2014reduced} it is stated that the code distance of a logical qubit stored using dislocations is 2L + O(1), where L is the separation between the dislocation twists.
This code distance assumed only physical X and Z errors are permitted \cite{hastings2019discuss}.
This short note shows that, when Y errors are allowed, the code distance reduces to L + O(1).
See \fig{error_chains}.
\end{abstract}

\vspace{-25pt}
\begin{figure}[h]
    \centering
    \resizebox{\linewidth}{!}{
    \includegraphics{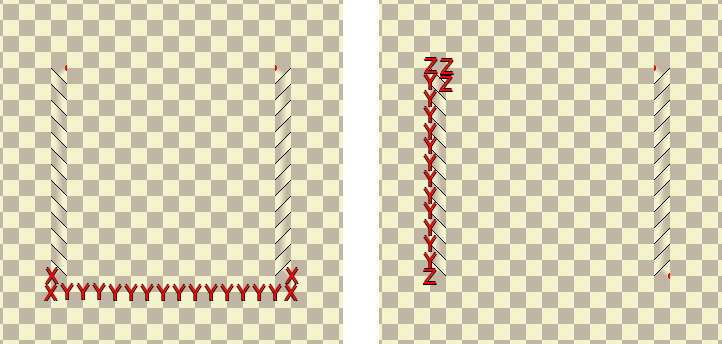}
    }
    \caption{
        Undetected error sets of size L + O(1) which anti-commute with the logical X and Z observables of the logical qubit stored using the dislocations.
        Light squares are Z stabilizers.
        Dark squares are X stabilizers.
        Data qubits are at the corners of the squares.
        Slanted squares are the dislocations, which have stabilizers with X on one side and Z on the other.
        At the end of each dislocation is a twist which has a stabilizer with X, Y, and Z observables.
        Red characters are a set of undetected errors.
        There are O(1) complications near each twist in order to arrange initially separate X and Z error chains into a Y error chain highway.
    }
    \label{fig:error_chains}
\end{figure}

\vspace{-16pt}
\section{Acknowledgements}

We thank Matthew Hastings and Austin Fowler for double-checking the existence of the short error chains and also for interesting (though inconclusive) discussion about whether the simulation results presented in the paper predict the existence of the short error chains.

\vspace{-4pt}
\bibliographystyle{unsrt}
\bibliography{refs}

\begin{thebibliography}{1}

\bibitem{hastings2014reduced}
Matthew~B Hastings and A~Geller.
\newblock Reduced space-time and time costs using dislocation codes and
  arbitrary ancillas.
\newblock {\em arXiv preprint arXiv:1408.3379}, 2014.

\bibitem{hastings2019discuss}
Matthew~B Hastings.
\newblock Private correspondence.
\newblock 2019.

\bibitem{lavasani2018low}
Ali Lavasani and Maissam Barkeshli.
\newblock Low overhead clifford gates from joint measurements in surface,
  color, and hyperbolic codes.
\newblock {\em Physical Review A}, 98(5):052319, 2018.

\bibitem{benbrowntwitter2019}
Ben Brown.
\newblock Twitter reply.
\newblock \url{https://twitter.com/BenBrow31417089/status/1179518287287177216}
  Accessed: 2019.

\end{thebibliography}

\appendix
\section{Comments}

Lavasani and Barkeshli have informed me of previous work in their 2018 paper \cite{lavasani2018low}.
The last paragraph of page 21 of that paper notes that the distance of dislocation qubits is less than $2L + O(1)$.

Ben Brown notes \cite{benbrowntwitter2019} that although the code distance is $L+O(1)$, the limitation on the directions Y errors can travel allows a packing that is twice as dense the one implied by \fig{error_chains}.

\end{document}